\documentclass{article}
\usepackage{spconf,amsmath,graphicx}
\usepackage{multirow}
\usepackage{subfig}
\usepackage{enumitem}
\usepackage{cite}

\usepackage[dvipsnames]{xcolor}

\title{A Real-Time Lyrics Alignment System Using Chroma and Phonetic Features for Classical Vocal Performance}
\name{Jiyun Park \qquad Sangeon Yong \qquad Taegyun Kwon \qquad Juhan Nam
\thanks{This work was supported by the National Research Foundation of Korea (NRF) grant funded by the Korea government (MSIT) (No. NRF-2023R1A2C3007605).}}

\address{Graduate School of Culture Technology, KAIST, Daejeon, South Korea}
\begin{document}
\ninept
\maketitle
\begin{abstract}
The goal of real-time lyrics alignment is to take live singing audio as input and to pinpoint the exact position within given lyrics on the fly.
The task can benefit real-world applications such as the automatic subtitling of live concerts or operas.
However, designing a real-time model poses a great challenge due to the constraints of only using past input and operating within a minimal latency.
Furthermore, due to the lack of datasets for real-time models for lyrics alignment, previous studies have mostly evaluated with private in-house datasets, resulting in a lack of standard evaluation methods.
This paper presents a real-time lyrics alignment system for classical vocal performances with two contributions.
First, we improve the lyrics alignment algorithm by finding an optimal combination of chromagram and phonetic posteriorgram (PPG) that capture melodic and phonetics features of the singing voice, respectively.
Second, we recast the Schubert Winterreise Dataset (SWD) which contains multiple performance renditions of the same pieces as an evaluation set for the real-time lyrics alignment.

\end{abstract}
\begin{keywords}
audio-to-lyrics alignment, real-time audio processing, score following, singing voice
\end{keywords}
\section{Introduction}
\label{sec:intro}

In classical vocal performances such as operas, surtitles are often displayed to help the audience understand the lyrics.
Since classical performances often involve tempo variations, short pauses, and differing pronunciations by singers, many venues hire human operators to control the lyrics displayed in tandem with the performance manually.
The goal of real-time lyrics alignment is to automate the operation by tracking the exact position of a song and matching it to lyrics on the fly. 
% This can benefit from the real-time lyrics alignment task. The goal of real-time lyrics alignment is to track the precise location of a song in real-time using pre-defined lyrics and live singing audio.

% ======== Related Works ==========
% 기존 audio-to-lyrics alignment 연구
Previous methods of audio-to-lyrics alignment mainly consist of two steps \cite{sharma2019automatic,stoller2019end,vaglio2020multilingual,gupta2020automatic,huang2022improving,ismirKruspe16}; (1) transforming the two different modalities, audio and text, into common linguistic or phonetic units as an intermediate representation, (2) aligning the two sequences to predict the timestamps of lyrics in different text units such as line, word, or character. The audio modality entails an acoustic model that classifies each audio frame into a set of textual or phoneme units. The advances of acoustic model have been largely borrowed from automatic speech recognition (ASR) \cite{sharma2019automatic,stoller2019end,vaglio2020multilingual,gupta2020automatic}. 
The text modality is often converted into phoneme sequences by a pronunciation dictionary such as CMUdict\footnote{http://www.speech.cs.cmu.edu/cgi-bin/cmudict} or grapheme-to-phoneme(g2p) tools \cite{huang2022improving, brazier2021line, durand2023contrastive}.  
Recently, pitch detection was additionally incorporated to improve the alignment performance \cite{huang2022improving}.
%\cite{huang2022improving} improved the alignment performance with a multitask learning approach that uses pitch detection in addition to phonemes. 
As to the alignment step, dynamic time warping (DTW) \cite{sharma2019automatic, brazier2021line} and Hidden Markov models (HMM) \cite{stoller2019end, vaglio2020multilingual, huang2022improving} have been widely used.  
%Regarding the alignment algorithm, while Viterbi-based forced alignment was mainly used \cite{stoller2019end, vaglio2020multilingual, huang2022improving}, algorithm based on dynamic time warping (DTW) \cite{sharma2019automatic, brazier2021line} was also widely used within MIR research.
While most previous studies have adopted source separation from the audio recordings in order to perform alignment on separated vocals without background music \cite{sharma2019automatic,vaglio2020multilingual}, recent studies have increasingly incorporated singing voices mixed with background music \cite{gupta2020automatic,demirel2021low}.
However, most previous audio-to-lyrics alignment methods are designed for offline use, such as displaying lyrics on a music streaming service, and cannot be applied directly in real-time scenarios.

% 실시간 연구
The methods for real-time settings have been widely studied in the domain of real-time music tracking, also called score following.
The main algorithms used in the score following include online versions of DTW \cite{dixon2005line, macrae2010accurate, arzt2015real} and HMM \cite{cont2009coupled, montecchio2011unified, otsuka2011real}.
For score following methods, using reference audio as a proxy score is common rather than directly aligning the performance audio with the score \cite{arzt2015real, dixon2005live}.
This can be achieved by pre-aligning the reference audio and synthesized audio from a deadpan MIDI extracted from a symbolic score \cite{arzt2015real}.
The most recent study dealing with real-time lyrics alignment \cite{brazier2021line} applied this score following approaches by utilizing reference performances.
Since \cite{brazier2021line} targeted for full opera tracking, especially focusing on \textit{recitative}, it adopted a phonetic posteriorgram (PPG) from mel-frequency cepstral coefficients (MFCC) as a feature, instead of chroma that is traditionally used in score following.
Additionally, the dataset used for evaluation in \cite{brazier2021line} is manually annotated data for full opera recordings, which is heavily labor-intensive and not publicly available.

% 우리 연구 본격 소개
In this study, we propose a full pipeline of the real-time lyric alignment system on the most general form of classical vocal performance, voice with piano accompaniment. In particular, we tackle this problem in a more score-following manner by utilizing reference (\textit{ref}) audio and symbolic score.
The system is able to capture both phonetics and melodic features of the singing voice by extracting PPG via our proposed acoustic model and chromagram.
The main contributions of the proposed system can be summarized as follows:

\begin{itemize}% [nosep]
    \item We align lyrics and audio via symbolic score without relying on manual annotations by leveraging the methods of score-following. %language-dependent pronunciation tools or annotating manually.
    \item We investigate an optimal combination of chromagram and PPG for a robust real-time singing alignment model. 
    \item We recast the Schubert Winterreise Dataset (SWD) to \textit{winterreise\_rt}, a subset dataset suitable for evaluating the real-time lyric alignment.
\end{itemize}
% Instead of relying on language-dependent tools or manually annotating the timing of each lyric unit of the reference audio, we fully utilize the symbolic scores and MIDI-rendered audio to extract the exact note-wise timings and lyrics.
We report alignment results on several benchmark tests for different feature combinations based on the \textit{winterreise\_rt} dataset.
We also share the script for constructing \textit{winterreise\_rt} from the original SWD dataset.
The demo video and additional information about the evaluation dataset are available on our companion website.
\footnote{https://laurenceyoon.github.io/real-time-lyrics-alignment/}

\begin{figure*}[t]
  \centering
  \centerline{\includegraphics[width=0.95\linewidth]{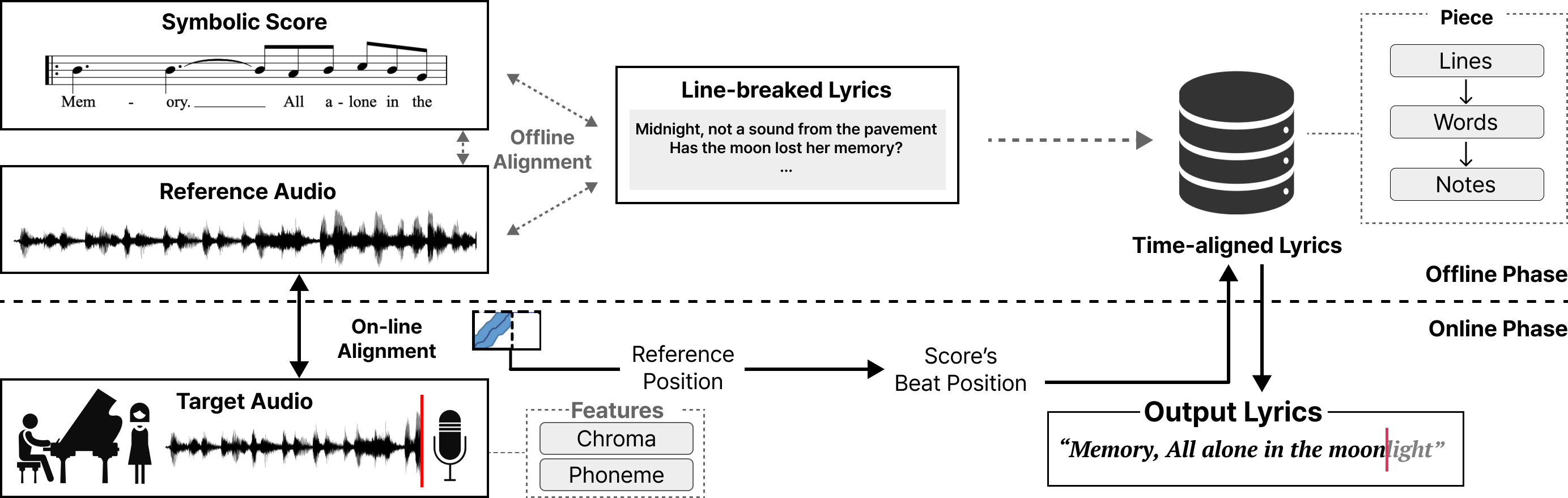}}  
\caption{An overview of the proposed real-time lyrics alignment system.}
\label{fig:overview}
\end{figure*}

\section{System Design}
\label{sec:system_design}

As illustrated in Figure \ref{fig:overview}, our proposed real-time lyrics alignment system consists of two phases: an offline phase which pre-aligns lyrics and reference audio, and a subsequent online phase during a live singing performance.
The core components of the system include an acoustic model and an online alignment algorithm. 

\subsection{Offline Phase}
\label{ssec:offline}

\begin{figure}[t]
  \centering
  \centerline{\includegraphics[width=\columnwidth]{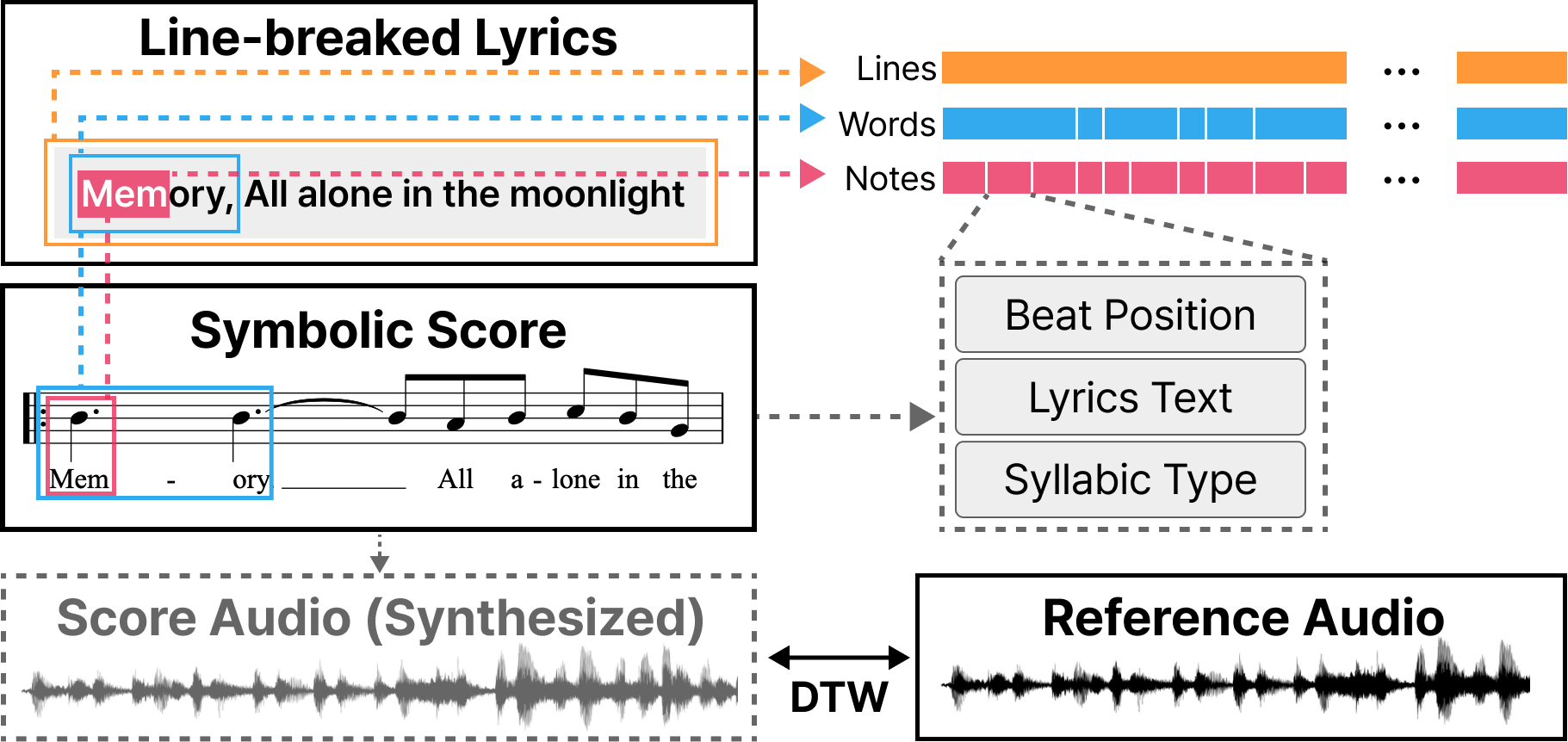}}
\caption{The pipeline of an offline phase aligning \textit{ref} audio and lyrics by extracting \textit{score} audio and annotation from the symbolic score.}
\label{fig:offline}
\end{figure}

The offline phase aims to obtain precise alignment between the \textit{ref} audio and lyrics so that the system can calculate the lyrical position only from \textit{ref} and \textit{target} audio in online phase, as illustrated in Fig.\ref{fig:offline}.  
The motivation for the design was to allow the system to automatically generate pseudo-labels from the symbolic score, even if there are no time-aligned lyrics labels for the \textit{ref} audio.
Specifically, we extract each onset timings of vocal notes in beat position, syllables of lyrics mapped to specific notes, and the syllabic type (\textit{start, end, or middle}) from the symbolic score as MusicXML format.

The \textit{ref} audio is DTW-aligned with the \textit{score} audio, which is the MIDI-synthesized audio from the symbolic score.
% Although we did not take advantage of the phonetic features of the lyrics when synthesizing the \textit{score} audio, we adopted an organ sound for the vocal part to produce a more sustained tone.
We used memory-restricted multi-scale DTW (MrMsDTW) \cite{pratzlich2016memory} and Chroma \& DLNCO (decaying local adaptive normalized chroma onset) for the feature for the offline alignment algorithm, provided by \texttt{synctoolbox}\cite{muller2021sync}.
Since the DLNCO is unsuitable for real-time settings because the framewise computation considers both past and future input simultaneously, we use it only for the offline phase.

Adapted from the four-level granularity (paragraph, line, word, note) of DALI \cite{meseguer2019dali}, the largest open-source time-aligned lyrics dataset, we construct our time-aligned lyrics with three-level of granularity: line, word, and note.

\subsection{Online Phase}
\label{ssec:online}

The online phase includes feature extraction of the \textit{ref} and \textit{target} audio on the fly and an online alignment algorithm between the two extracted features.

Given two sequences $X := (x_1, \ldots, x_N)$ with length $N$,  $Y := (y_1, \ldots, y_M)$ with length $M$ and distance function $d(x_n, y_m)$, the standard Dynamic Time Warping (DTW) algorithm aims to find an optimal warping path $W := (w_1, ..., w_K), w_k = (n, m) $ such that the sum of the computed distance costs $\sum_{W}{d(x_n,y_m)}$ is minimized.\footnote{For detailed definition of the algorithm, please refer \cite{dixon2005line}}
However, the standard DTW cannot be directly applied to real-time due to two limitations; (1) inherent \textit{backtracking} procedure requires the entire cost matrix beforehand, and (2) it is computationally intensive with quadratic time complexity.
As Online Dynamic Time Warping (OLTW) achieves linear time complexity and optimizes for real-time by incremental solution, we reproduced the algorithm based on \cite{dixon2005line} with different configurations suitable for the singing model.
We set a sample rate of 16kHz, frame rate of 25, and OLTW window size \cite{dixon2005live} of 3 seconds.

To process audio in real-time, we implemented a stream processor that enqueues the streaming audio in chunks and extracts features of the \textit{target} audio. 
The size of the audio buffer corresponds to 160ms.
While the OLTW algorithm is running, the \textit{ref} position is transferred into the \textit{score} position via linear interpolation, followed by calculating the score's beat position.

\subsection{Acoustic Model}
\label{ssec:acoustic_model}

The main task of an acoustic model is to predict the probability of a sequence of phonetic or linguistic units given a sequence of audio features.
In our system, the acoustic model runs in the online phase to compute the phonetic similarity for the \textit{ref} and \textit{target} audio.
In previous models \cite{vaglio2020multilingual,ismirKruspe16,brazier2021line}, the audio was converted into phonemes by acoustic models, and the lyrics were also converted into the same phoneme set to run forced alignment between two modalities.
In our system, however, we only use the acoustic model that converts each of \textit{ref} and \textit{target} audio into phonetic posteriorgram (PPG) in order to be used as features for real-time audio-to-audio alignment.
The audio-to-lyrics alignment is indirectly preprocessed at the offline phase via a symbolic score as described in Section \ref{ssec:offline}.

\begin{figure}[t]
  \centering
  \centerline{\includegraphics[width=\columnwidth]{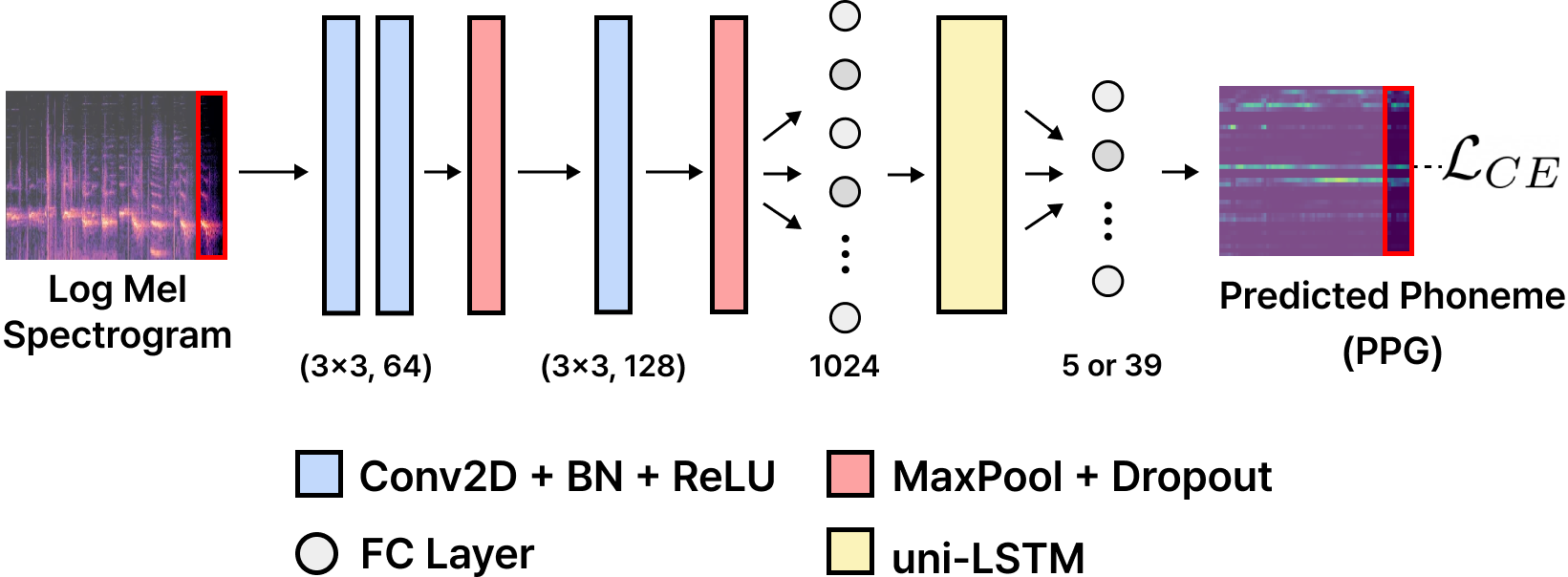}}  
\caption{The network architecture of the proposed acoustic model}
\label{fig:classifier}
\end{figure}

Using the CRNN as the backbone architecture, our proposed acoustic model consists of a single CRNN network with a dense layer that takes log-scaled mel-spectrogram and outputs a PPG matrix as illustrated in Fig.\ref{fig:classifier}.
The overall architecture is taken from the framewise phoneme classifier in \cite{yong2023phoneme}, but modified to be suitable for real-time.
The \textit{ConvNet} architecture, proposed in \cite{kelz2016potential}, was used in the CNN part followed by an uni-directional LSTM layer with 1024 size as the RNN part.
The last fully-connected layer outputs a PPG matrix with a target size of $N_{phone} \times N_{frame}$.
As we used a dataset with time-aligned phoneme labels for training, we employed the cross-entropy loss.
The model takes as input a 66-bin mel-spectrogram extracted from an audio signal, hop size of 640, FFT size of 1280 with a sampling rate of 16kHz.

\vspace{-5pt}
\section{Experiments}
\label{sec:experiments}

\vspace{-5pt}
\subsection{Datasets}
\label{ssec:dataset}

% To train an acoustic model, the connectionist temporal classification (CTC) loss has been commonly used \cite{vaglio2020multilingual, huang2022improving,brazier2021line}. 
% That's because obtaining time-aligned datasets between lyrics and phonemes is a great challenge and CTC-based training can achieve alignment using only a weakly annotated, yet large-sized dataset.

We used TIMIT \cite{garofolo1993timit} to train our proposed acoustic model, which has already shown robustness in singing voice alignment \cite{ismirKruspe16,yong2018singing} and singing transcription \cite{yong2023phoneme}. 
Although comprised of English-only speech audio, the model can also be applied to multilingual scenarios as it only compares the phonetic similarity between two audio sequences.
We use three types of collapsed sets: (1) 39 phone sets collapsed from the original 61 TIMIT labels proposed by \cite{lee1989speaker}, denoted as `Phoneme39', (2) 14 viseme sets, which are a set of phonemes with similar visual articulations, from \cite{saenko2004articulatory}, denoted as `Viseme14', (3) and 5 broad phonetic groups as \textit{vowels, stops, fricatives, nasals}, and \textit{silences} \cite{deekshitha2018novel}, denoted as `Phoneme5'.
We used the cross-entropy loss function in framewise, as the TIMIT corpus provides a phoneme-level of time-aligned annotation.

Evaluating real-time lyrics alignment for singing voice is challenging because it requires a dataset with multiple performance song renditions with each label of time-aligned lyrics. 
To the best of our knowledge, there was no standard public dataset to benchmark real-time lyrics alignment models. 
We reconstructed Schubert Winterreise Dataset(SWD) \cite{weiss2021schubert} into \textit{winterreise\_rt}, a subset that enables the benchmark evaluation of real-time lyrics alignment models.
The SWD dataset is a collection of resources of Schubert's song cycle for voice and piano `Winterreise', one of the most-performed pieces within the classical music repertoire \cite{weiss2021schubert}.
Out of the two publicly available versions of the recording, we used \textit{HU33} as the \textit{ref}, \textit{SC06} as the \textit{target}, and the audio rendered from MIDI as the \textit{score}.
The \textit{target} audio has been transposed to match the key of the \textit{ref} audio, and only voice note annotations were retained from the original note annotations.
Detailed information about the dataset is available at this link \footnote{https://github.com/laurenceyoon/winterreise\_rt}.

\vspace{-5pt}
\subsection{Evaluation Metrics}
\label{ssec:evaluation_metrics}

From the classic evaluation metrics from the score following \cite{cont2007evaluation} and the international MIREX\footnote{ http://www.music-ir.org/mirex.} challenge for audio-to-lyrics alignment, we use these three evaluation metrics to evaluate our model. 
\begin{itemize}[nosep]
    \item \textbf{Average Absolute Error (AAE)}: the mean absolute time error of voice note-level timings.
    \item \textbf{Median Absolute Error (MAE)}: the median absolute time error of voice note-level timings.
    \item \textbf{Percentage of Correct Onsets (PCO)}: the ratio of correctly estimated word onset times within a tolerance threshold $\theta$. (tolerance $\theta$: 200ms, 300ms, 500ms, 1000ms)
\end{itemize}

\subsection{Feature Selection \& Hyper-parameters}
\label{ssec:feature}

The audio features used in the experiment include chromagram computed from an STFT, a 66-bin mel-spectrogram, 13- and 5-dimensional MFCCs from a standard library\footnote{https://librosa.org}, and the phonetic posteriorgram (PPG) extracted from our acoustic model in Section \ref{ssec:acoustic_model}.
For the PPG-based feature, we used three types of feature: `Phoneme39', `Viseme14' and `Phoneme5' as mentioned in Section \ref{ssec:dataset}.
We performed a grid search for the optimal normalization(L2, L-inf), scaling(softmax, log1p), and distance metrics(euclidean, cosine) for each feature type. 
The \textit{log1p} function is $\log1p(x) = \frac{\log(a \cdot x + 1)}{b}$ where $x$ denotes a feature vector input. We use $a=5, b=4$ for the scaling parameter, resulting in the value range compressed below $0.5$.
Based on our empirical results, we applied L-inf norm + \textit{log1p} with Euclidean distance for chroma, L-inf norm + \textit{log1p} with cosine distance for mel-spectrogram, and softmax + \textit{log1p} with cosine distance for PPG.
% When extracting features, the usual centered frame method with front-to-back padding is not applied because there should not be any unnecessary paddings in the middle of the signal when the signal is chunked into blocks.

\begin{table*}[h]
\resizebox{\textwidth}{!}{%
\def\arraystretch{1.05}
\begin{tabular}{c|c|c|c|c|cccc}
\hline\hline
\multirow{2}{*}{Phase} &
  \multirow{2}{*}{Feature Type} &
  \multirow{2}{*}{AAE (ms) ↓} &
  \multirow{2}{*}{MAE (ms) ↓} &
  \multirow{2}{*}{Std. (ms) ↓} &
  \multicolumn{4}{c}{Percentage of Correct Onsets(PCO) (\%)} \\ \cline{6-9} 
    &&&&& \textless 200ms ↑ & \textless 300ms ↑ & \textless 500ms ↑ & \textless 1000ms ↑ \\ \hline
\textbf{Offline} &
  \textbf{Chroma \& DLNCO} &
  \textbf{302} &
  \textbf{137} &
  \textbf{733} &
  \textbf{67.71} &
  \textbf{83.35} &
  \textbf{93.31} &
  \textbf{98.1} \\ \hline\hline
\multirow{5}{*}{\textbf{Online}} & Chroma             & 1,366 & 1,098 & 1,086 & 58.99          & 69.42          & 79.19  & 85.92\\
                                 & Mel spectrogram    & 1,222 & 1,062 & 692   & \textbf{66.69} & \textbf{75.75} & 82.22  & 86.71\\
                                 & Phoneme39          & 2,637 & 1,924 & 2,294 & 31.94          & 37.25          & 43.39  & 52.09\\
                                 \cline{2-9}
                                 & Chroma + MFCC (13) & 1259  & 950   & 934   & 59.54          & 69.56          & 79.06  & 85.96\\
                                 & Chroma + MFCC (5)  & 453   & 144   & 766   & 62.85          & 73.13          & 83.46  & 90.56\\
                                 \cline{2-9}
                                 & Chroma + Phoneme39 & 1,010 & 511   & 1,317 & 63.96          & 73.32          & 82.13  & 88.04\\
                                 & Chroma + Viseme14  & 1,007 & 735   & 1,010 & 63.62          & 73.65          & 83.65  & 89.63\\ 
                                 &
  \textbf{Chroma + Phoneme5} &
  \textbf{376} &
  \textbf{136} &
  \textbf{630} &
  64.75 &
  75.49 &
  \textbf{85.82} &
  \textbf{92.26} \\ 
  \hline\hline
\end{tabular}%
}
\caption{Results of offline \& online alignment on \textit{winterreise\_rt} dataset. Offline alignment results are evaluated with each ground-truth of \textit{score} and \textit{ref}, and online alignment results are evaluated with each ground-truth of \textit{ref} and \textit{target} using each feature type (all values are averaged over 24 songs with voice note-level evaluation).}
\label{table:result}
\end{table*}

\section{Results \& Discussions}
\label{sec:results}

% Result 개요 + offline result 분석
% MFCC로만 테스트한 건 결과가 안좋았음 추가하기
As shown in Table \ref{table:result}, we report the results of alignment accuracy for offline and online phases, respectively.
Since the offline results are used as pseudo-ground truth for the online alignment when the system is operated, the offline alignment result may also affect the overall alignment result.
For rigorous verification of each phase, we evaluated each alignment results from the extracted the ground-truth for each of \textit{score}, \textit{ref}, and \textit{target}.
All values are averaged over 24 songs and evaluated at the voice note-level.
The results of the online alignment are reported based on the optimal hyper-parameters for each feature type's normalization, scaling, and distance metric mentioned in Section \ref{ssec:feature}.

For the offline results, the AAE and MAE are calculated to be 302 ms and 137 ms, with the highest PCO ratio for all phases and features, which is well within the acceptable range that listeners typically perceive.
Interestingly, the \textit{ref} audio is aligned fairly well with the \textit{score} audio despite synthesizing with only instrumental sounds and leaving out the lyric information, although may not be comparable to the state-of-the-art models \cite{huang2022improving, durand2023contrastive}.

% online result 분석
According to the online alignment results from Table \ref{table:result}, our model that uses Chroma \& Phoneme5 outperforms all other feature types in AAE (376 ms) and MAE (136 ms), and is even comparable to the offline results.
We observe that the Chroma-only, common in traditional score-following models, did not work well for singing.
The mel-spectrogram outperformed chroma and even had the highest PCO at 0.2 and 0.3 seconds, presumably due to its advantages in capturing both melodic and phonetic characteristics.
However, its performance lags behind a PPG-included model that solely extracts phonetic characteristics from our acoustic model.
Although Phoneme39 underperforms on its own, it surpasses the mel-spectrogram when paired with chroma.
Outcomes using the standard 13 MFCC were unremarkable, whereas the significant reduction to 5 dimensions yielded notably better results.
Its performance peaks when MFCCs are replaced with the PPG from Phoneme5 model.
% \footnote{We do not report Phoneme5-only model because it performed significantly worse than the Phoneme39-only model.}

\textbf{Both melodic and phonetic features are required for the real-time lyrics alignment for classical vocal performance.}
The results show that the model combined with chroma and phoneme features significantly improves the performance.
This reflects the importance of characteristics of classical vocal performance, which emphasizes both rich melodic expression and pronunciation of lyrics.
% In classical singing, singers are often responsible for leading an accompanist and the timing between the singer and accompanist may vary.
% The model can be more sensitive to the singer's pronunciation by using phonemes.

\textbf{Chroma are relatively more significant than phonemes.}
% At least for live classical vocal performance with a vocal and piano, we see that chroma is relatively more significant than phonemes and that phonemes are still necessary but may play a secondary role.
Our experiments showed that the model using Chroma + Phoneme5, which has a relatively higher proportion of Chroma in the feature matrix and a smaller number of phone sets, outperformed the models with a larger set of classifications.
Also, we found that leaving the chroma value between [0, 1] but compressing the range of phoneme values below 0.5 with a \textit{log1p} function contributed to better alignment performance.
% Furthermore, reducing phonetic information with similar characteristics was more effective in improving alignment than the phonemes with larger phone sets.
% This suggests that for computing the similarity of two classical singing audio features in real-time, a small number of phone sets is sufficient to capture only distinctive phonetic changes, and pitch is a more important variable.
The predominance of pitch over phonetic information in the real-time computation of audio feature similarity is underscored in classical vocal music, attributed to its rich melodic intricacies and the inherent ambiguity in enunciation.
As classical singing has been considered by far the least intelligible genre among twelve singing genres \cite{condit2015catching}, our results reflect that it is more important to detect the melodic variations than to detect the micro phonetic deviation for the singers.

\begin{figure}[t]
  \centering
  \centerline{\includegraphics[width=0.97\columnwidth]{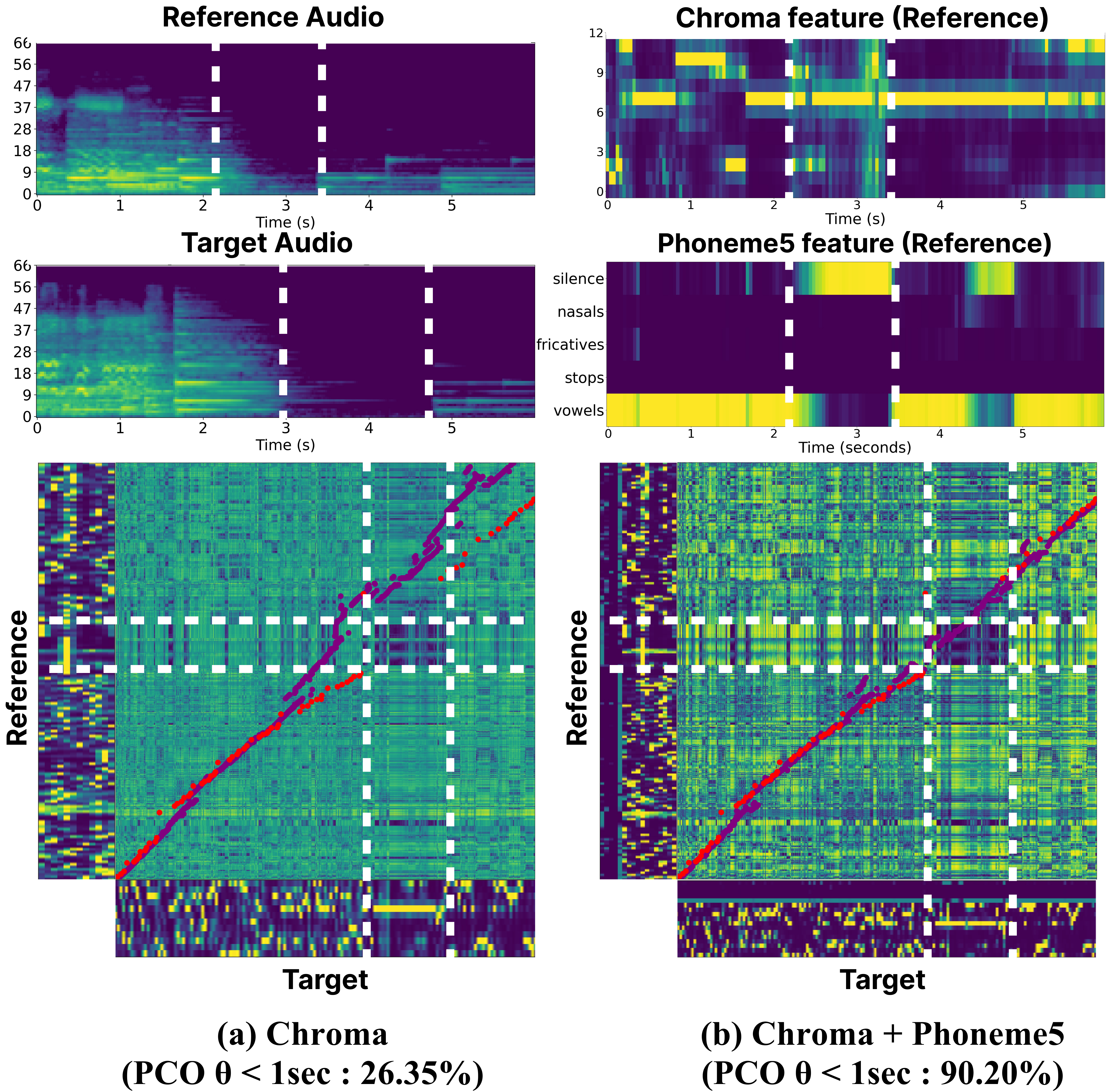}}  
  \vspace{-5pt}
\caption{Warping path results with and without phoneme features of `No. 11, Frühlingstraum'.
The red dots represent ground-truth pairs for voice notes, the purple dots with the warping path results of each model, and the white line with the silence part.}
\label{fig:discussion}
\end{figure}

\textbf{Phonemes are robust against silence \& sudden voice onset.}
As we further investigated, we found that adding phonemes makes it particularly robust against prolonged silence with a sudden voice onset, which is especially important in live performances.
Figure \ref{fig:discussion} compares the result of with and without phonemes through the ground-truth labels (red) and warping path (purple).
Analyzing the two audio recordings from 00:52 to 01:02, where the warping path deviation starts, we found that the voice onset of the \textit{target} audio enters after a prolonged period of silence.
In classical singing, singers are often responsible for leading an accompanist and the note timing between the singer and accompanist may vary.
Since the phoneme features capture the silence and fricatives separately, 
The model can be more sensitive to the singer's pronunciation by using phonemes.

\section{Conclusion}
\label{sec:conclusion}

\vspace{-5pt}
This study presented a full pipeline of the real-time lyrics alignment system for classical vocal performances using a CRNN-based acoustic model and chromagram.
% We propose a CRNN-based acoustic model that can be applied to online classical singing alignment.
Our system achieves the automated temporal alignment between lyrics and singing voice by taking advantage of the score following methods.
We utilize the public SWD dataset to evaluate the real-time lyrics alignment task for the first time, recasting it into \textit{winterreise\_rt}.
The result shows that integrating chroma and phonetic features in the model substantially improves the alignment results, highlighting the rich melodic expression with lyrical content.
% We plan to consider jointly adopting linguistic tools, such as G2P tools, with the symbolic score on the offline phase to improve the offline alignment performance.
% Future work will also need to improve the acoustic model more adapted to real-time polyphonic singing since the training data of our acoustic model was speech-only.
% Furthermore, the evaluation was limited to a small, male-only classical dataset due to the need for real-time alignment. due lack of real-time lyrics alignment 
% Our future work involves improving the acoustic models that are optimized for the characteristics of real-time singing, such as instrument-only interludes.
As a future work, we aim to enhance the acoustic model to more accurately represent the characteristics of real-time polyphonic singing, particularly to be robust in the presence of accompaniment, thus surpassing the limitations of models trained solely on speech data.
% For future work, we plan to gather a larger dataset based on the structure of \textit{winterreise\_rt} to improve the system and investigate feature differences across genres, languages, and voice types.
% Also, our additional empirical experiments found that the proposed model is robust to other languages, such as Latin and Korean.
% As future work, we plan to conduct cross-linguistic experiments with a larger dataset to validate its scalability to multiple languages.
% Multilingual scenarios로 확장될 가능성이 있다
% Also, the proposed system requires a specific format of symbolic score, a MusicXML which is very cumbersome to process.
% As we are aware that most existing lyrics datasets do not contain MusicXML, we will consider it as future work to incorporate existing lyrics datasets for evaluation.

\vfill\pagebreak

\bibliographystyle{IEEEbib}
\bibliography{refs}

\end{document}